\documentclass[12pt,onecolumn]{IEEEtran}

\usepackage{tikz}
\usepackage{lineno,hyperref}
\usepackage{amsmath,amscd,amssymb,amsgen,amsfonts,amsbsy}
\usepackage{breqn}
\usetikzlibrary{automata,arrows,positioning,calc}
\newtheorem{theorem}{Theorem}
\newtheorem{lemma}{Lemma}
\newtheorem{remark}{Remark}
\newtheorem{definition}{Definition}

\renewcommand{\baselinestretch}{0.97}

\title{Multi-armed Bandits with Constrained Arms and Hidden States
  \thanks{The work of Varun Mehta and Kesav Kaza was done in SPANN Lab
    at IIT Bombay. The work of Rahul Meshram was carried out in the
    Bharti Centre for Communications at IIT Bombay.}}

\author{Varun Mehta, Rahul Meshram, Kesav Kaza and
  S.~N.~Merchant \\ 
 Department of Electrical Engineering,
  \\ Indian Institute of Technology Bombay, India
}

\begin{document}
\maketitle



%
%

\begin{abstract}  
The problem of rested and restless multi-armed bandits with
constrained availability of arms is considered. The states of arms
evolve in Markovian manner and the exact states are hidden from the
decision maker.  First, some structural results on value functions are
claimed. Following these results, the optimal policy turns out to be a
\textit{threshold policy}. Further, \textit{indexability} of rested
bandits is established and index formula is derived. The performance
of index policy is illustrated and compared with myopic policy using
numerical examples.
\end{abstract}




\section{Introduction}
Multi-armed bandits are among commonly used models for solving
sequential decision making problems, \cite{Gittins79,Gittins11}. In
the multi-armed bandit problem, there are $N$ arms and each arm can be
in one of a finite set of states. The decision maker plays $M$ arms,
$(M < N)$ at every time instant and collects rewards from the played
arms. Reward from each played arm depends on the state of that
arm. The state of an arm changes according to a stochastic process
associated with that arm. The decision maker's aim is to maximize the
long-term expected discounted reward. The state evolution may be
action dependent and based on that there are two types of bandits,
rested and restless bandits. In a rested bandit, the state evolves
only for the arm which is played while states of other arms do not
change. For a restless bandit, the states of all arms evolve even when
they are not played.  In this setting, each arm can be considered as a
Markov decision process (MDP) with finite states and two actions (play
or not to play) in each state.  As a model choice, states may assumed
to be either observable by decision maker or hidden to it. Now, the
multi-armed bandit problem can be looked as a set of MDPs coupled
together with constraints.

A rested multi-armed bandit problem was first introduced in the
seminal work of \cite{Gittins79}, where the author proposed an index
based policy.  In such policies, state of each arm is mapped to an
index, i.e., real valued number. At each time instant arms with the
highest indices are played. This policy is known as Gittins index
policy. Later, a generalization of the rested multi-armed bandit
problem was devised in \cite{Whittle88}, where a restless multi-armed
bandit was introduced and again index based policy proposed. The index
policy for restless bandits is now referred to as Whittle index
policy.

Recently, restless bandits have been studied when state of the arms
are not observable but feedback signal is observable. The decision
maker estimates the state from this feedback. This is called the
hidden Markov bandit. For a hidden Markov bandit, each arm can be
modeled using partially observable Markov decision processes (POMDP).
An index policy for hidden Markov rested multi-armed bandit is
suggested in \cite{Krishnamurthy09}. Further, extension of this to
hidden Markov restless bandit is analysed in work of
\cite{Meshram16,Meshram17,Borkar17,LiuZhao10}.

To use index policy in rested and restless bandit, an approach is to
first consider the single-armed bandit problem and show that the
optimal policy is of a threshold type. Using this result one can show
that arm is indexable and later index can be derived. While analyzing
a single-armed bandit model, structural results of POMDP can be used
for hidden Markov bandits. Some structural results for POMDP have been
extensively studied in \cite{Lovejoy87,Lovejoy87a,Albright79,White79}.

All of the above works on bandits assume that every arm is available
for decision maker at each time instant to play. The decision maker
determines whether to play or not play the arms using index
policy. But this may not be feasible in some scenarios. For example,
in a machine-repair problem one may not able to schedule a task on
some of the machines due to machine breakdown. Such consideration has
been made in \cite{Dayanik02}. In queuing systems, the controller may
not be able to schedule jobs to some servers due to server breakdown,
\cite{Martin05, Glazebrook07}. In these examples a machine or server
is available to the decision maker intermittently. In this work, we
consider rested and restless bandits with arm availability constraints
where arms may not be available to play at some time instants and
these are called as constrained bandits. It is a generalization of the
classical rested and restless multi-armed bandit problems. Usually
when arm is not available, we consider a substitute arm which yield
low reward compare to the arm when it is available.

In constrained bandits \cite{Dayanik02,Martin05,Glazebrook07}, each
state is defined as a pair $(X(t),Y(t)),$ where $X(t)$ represents the
state of arm and $Y(t)$ represents availability of an arm at time
$t$. Time is discretized in \cite{Dayanik02} while it is continuous in
the models of \cite{Martin05,Glazebrook07}. The state $(X(t),Y(t))$ is
assumed to be observable. Under some assumptions on model parameters
the index policy is analyzed in
\cite{Dayanik02,Martin05,Glazebrook07}.  In this paper we consider a
hidden Markov model, where state $X(t)$ of the arm is not observable
but the availability of the arm is observable.


The paper is organized as follows. In next section, we describe the
hidden Markov model for multi-armed bandit with constraints.  We later
consider single armed bandit problem in Section~\ref{sec:sab}. We
analyze structural results for single-armed bandit in
Section~\ref{sec:struct-results}. 
Section~\ref{sec:sabindex} we compute the index for hidden Markov
rested bandit with availability constraints on arm.  We also
illustrate the performance of the index policy and compare it with
that of myopic policy in Section~\ref{sec:nume_result}.  We finally
conclude in Section~\ref{disc} and discuss some of open issues.

\section{Preliminaries and Model Description}
Consider a multi-armed bandit with $N$ independent arms.  Each arm can
be in one of two states, $0$ and $1,$. The system is time slotted and
it is indexed by $t.$ Let $X_n(t)$ denote the state of arm $n$ at
beginning of time slot $t,$ $X_n(t) \in \{0,1\}.$ Each arm has
availability constraints i.e. it is intermittently available. Let
$Y_n(t) \in \{ 0,1\}$ represent the availability of arm $n$ in time
slot $t$ and
 \begin{eqnarray*}
 Y_n(t) = 
 \begin{cases}
 1 & \mbox{if arm $n$ is available,} \\
 0 & \mbox{if arm $n$ is not available.}
 \end{cases}
\end{eqnarray*}  
When arm $n$ is not available in slot $t$ we will assume that , the
arm $n$ is replaced by substitute arm which yield low reward after
play.
 $A_n(t) \in \{0,1 \}$ is the action in slot $t$ with the
following interpretation.	
\begin{eqnarray*}
  A_n(t) = \begin{cases}
    1 & \mbox{if arm $n$ is played in slot $t$,}\\
    0 & \mbox{otherwise.} 
  \end{cases}
\end{eqnarray*}
Exactly one arm is to be played in each time slot. Arm $n$ changes
state at the end of time slot $t$ according to transition
probabilities that depend on $A_n(t),$ $Y_n(t)$ and it is defined as
follows.
{
\small{
\begin{eqnarray*}
\Pr \{ {X_n}(t + 1) = j~|~{X_n}(t) = i,{Y_n}(t) = y,{A_n}(t) = a\}= P_{ij}^n(y,a)
\end{eqnarray*}
}}
%

In every slot $t,$ a binary signal $Z_n^{y}(t)$ is observed from the
arm $n$ that is played. There is no observation from the arms that are
not played.  Thus
 \begin{equation*}
 Z_n^{y}(t) = 
 \begin{cases}
 1 & \mbox{play of arm $n$ is successful} \\
 0 & \mbox{otherwise.}
 \end{cases}
 \end{equation*}

Let $\rho_{n}(i,y)$ be the probability of success given that arm $n$
is played $A_n(t) =1,$ and $X_n(t) = i,$ $Y_n(t) = y$. We assume $
\rho_{n}(0,y) < \rho_{n}(1,y)$ for $y \in \{0,1\}.$
\begin{equation*}
 \Pr{\left(Z_n^{y}(t) = 1~|~X_n(t) = i, Y_n(t) = y, A_n(t) =1 \right) } = \rho_n(i,y).
\end{equation*}
Also, $R_n^{a}(i,y)$ is the reward obtained from playing arm $n$
given that, $X_n(t) = i,$ $Y_n(t) = y,$ $A_n(t) =a.$ Let
\begin{eqnarray*}
 & R_n^{1}(i,1)=r_{n,i},  \hspace{0.2in} R_n^{1}(i,0) = \eta_{n,i} \\
 & R_n^{0}(i,1) = 0, \hspace{0.2in}  R_n^{0}(i,0) = 0.
\end{eqnarray*}
Further, we will suppose that $0 \leq \eta_{n,0} < r_{n,0} <
\eta_{n,1} < r_{n,1} \leq 1$ for all $n.$ 
\begin{remark}\hspace{2cm} 
\begin{itemize}
\item The observation variable $Z_n^y(t)$ may have different meanings in
 different applications.  In communication systems, $Z_n^y(t)=1$ may
 mean an acknowledgement (ACK) of a successful transmission over a
 given link, \cite{Meshram16}. For a recommendation system, it may
 correspond to click or like by the user over a recommended item, see
 \cite{Meshram17}.
 \item Notice that $\eta_{n,i} \neq 0$; this means there is a non-zero
   reward for playing an arm even when it is not available. This
   captures application scenarios where broken (not available) arms
   can be repaired by playing them and paying a penalty from the
   reward.
 \end{itemize}
\end{remark}
The decision maker cannot directly observe states of the arms, and
hence it does not know the states at the beginning of each time
slot. But decision maker knows the probability of availability
$\theta_n^a(i,y)$ of arm $n$, at the beginning of next time slot
$t+1$; it is as follows
{\small{
\begin{equation*}
\theta_n^a(i,y) = \Pr{ \left({Y_n}(t+1) = 1|{X _n}(t) = i ,{Y_n}(t) = y,A_n(t) = a \right) }.
\end{equation*}
}}
However, the decision maker maintains a belief $\pi_n(t)$ about the
state of arm $n.$ It is the probability that the arm is in state $0$
given all past availability, actions, observations. This is given as
follows.
\begin{eqnarray*}
\pi_n(t) = \Pr{\left(  X_n(t) = 0~|~\left( Y_n(s) = y_s,A_{n}(s) ,Z_{n}^{y_s}(s)   \right)_{s=1}^{t-1} \right)}.
\end{eqnarray*}
Let $H_t$ denote the history,
\begin{eqnarray*}
H_t := \left(Y_n(s) = y_s,A_{n}(s) ,Z_{n}^{y_s}(s) \right)_{1 \leq n \leq N, 1 \leq s < t}.
\end{eqnarray*}

We can describe the state of arm $n$ at time $t$ by $S_n(t) =
(\pi_n(t), Y_n(t)) \in [0,1] \times \{0,1\}.$ $(S_1(t), \cdots
S_N(t))$ is the state information of the arms at the beginning of time
slot $t.$ Further, we can rewrite $\theta_n^a(i,y)$ as function of
$\pi$ in following form.
{
\small{
\begin{equation*}
\theta_n^a(\pi ,y) = \Pr ({Y_n}(t + 1) = 1|{\pi _n}(t) = \pi ,{Y_n}(t) = y,A_n(t) = a).
\end{equation*}
}
}
Hence the expected reward from playing arm $n$ at time $t$ given that
$Y_n(t) = y$ is
\begin{equation*}
\widetilde{R}_n^{1}(\pi_n(t),y) = \pi_n(t) R_n^{1}(0,y) + (1-\pi_n(t) R_n^{1}(1,y).
\end{equation*}
In each slot, exactly one arm is to played. Let $\phi(t)$ is the
policy by the decision maker such that $\phi(t): H_t \rightarrow
\{1, \cdots,N\}$ maps the history to one of the arm at slot $t.$ Let
\begin{eqnarray*}
A_n^{\phi}(t) = 
\begin{cases}
1 & \mbox{if $\phi(t) = n,$ } \\
0 & \mbox{if $\phi(t) \neq n.$}
\end{cases}
\end{eqnarray*} 
We are now ready to define the infinite horizon discounted reward
under policy $\phi$ for initial state information $(\underline{\pi},
\underline{y}),$ $\underline{\pi}=(\pi_1(1), \cdots, \pi_N(1))$ and
$\underline{y} = (y_1(1), \cdots, y_N(1)).$ It is given by
\begin{eqnarray}
V_{\phi}(\underline{\pi}, \underline{y}) = 
\mathrm{E}^{\phi}\left({\sum_{t=1}^{\infty} \beta^{t-1} 
\left[ \sum_{n=1}^{N} 
A_n^{\phi}(t) \widetilde{R}_n^{1}(\pi_n(t), Y_n(t) )
\right]
}\right). 
\label{eqn:opt1}
\end{eqnarray}  
Here, $\beta$ is discount parameter, $0<\beta < 1.$ The goal is to
find a policy $\phi$ that maximizes $V_{\phi}(\underline{\pi},
\underline{y})$ for given $ \underline{\pi} \in [0,1]^N,$
$\underline{y} \in \{0,1\}^N.$ The optimization problem
\eqref{eqn:opt1} is a multi-armed bandit problem with availability
constraints. This is generalized version of multi-armed bandits, where
it has partially observable states and availability constraints. In
general, this problem is known to be
PSPACE-hard,\cite{Papadimitriou99}. Index based policies are developed
in \cite{Gittins11,Whittle88} for rested and restless multi-armed
bandits. To study such index policies, a Lagrangian relaxed version of
problem \eqref{eqn:opt1} is analysed. In this relaxed problem,
complexity of problem reduced as it separates the solving one
multi-armed bandit problem to $N$ single-armed bandit problems. Thus
it reduces to calculating the index for each arm separately. The arm
with highest index is played in each time slot.

We next analyze the single-armed bandit problem in next section.

\section{Single-armed bandit problem}
\label{sec:sab}
For notational convenience, we will drop the subscript $n$, i.e., the
sequence number of the arm. As a widely used method for solving the
single arm bandit problem, a subsidy $w$ is assigned for not playing
the arm \cite{ Whittle88}.  In that case, optimization problem
\eqref{eqn:opt1} can be rewritten as follows.

{\small{
\begin{eqnarray}
V_{\phi}(\pi, y) = 
\mathrm{E}^{\phi}\left(\sum_{t=1}^{\infty} \beta^{t-1} 
\left[  
A^{\phi}(t) \widetilde{R}^{1}(\pi(t), Y(t) ) 
+w(1-A^{\phi}(t))\right]
\right), 
\label{eqn:opt2}
\end{eqnarray}
} } 
where action $A(t)$ under policy $\phi$ is
\begin{eqnarray*}
A^{\phi}(t) = 
\begin{cases}
1 & \mbox{if $\phi(t) = 1,$ } \\
0 & \mbox{if $\phi(t) =0.$}
\end{cases}
\end{eqnarray*} 
The objective is to find a policy $\phi$ that maximizes 
$V_{\phi}(\pi,y).$

Recall that the state evolution of arms may be action dependent. Based
on this, we can have two different types of bandits, rested and
restless single-armed bandit.  In rested single-armed bandit, state
evolves for the arm that is played and state of other arms do not
change.  For restless bandit model, state of all arms changes at each
time slot.

To simplify the model further, we assume that $P_{00}(y,a)= \mu _0$
and $P_{10}(y,a)= \mu _1$ for $a,y \in \{0,1\}.$\footnote{But in
  general, transition probabilities for available and unavailable arms
  could be different.}  We will also assume that $\rho(i,1) =
\rho(i,0) = r_i,$ $i \in \{0,1\}.$ Recall that $\pi (t)= \Pr (X(t)=0 |
H_t)$ and using Bayes rule, we can obtain the belief $\pi (t+1)$ as
follows.

{\small{
\begin{eqnarray*}
\pi(t+1) = 
\begin{cases}
\gamma_{z,y}(\pi(t)) & \mbox{if $A(t) =1,$ $Y(t) = y,$ and $Z^{y}(t) = z,$ } \\
\Gamma_{y}(\pi(t)) & \mbox{if $A(t) = 0,$ and $Y(t) = y.$}
\end{cases}
\end{eqnarray*}
}}
Here,
\begin{enumerate}
\item If $A(t)=1,$ i.e., arm is played and $Y(t)=1, Z^1(t)=1$ then 
\begin{displaymath}
 \gamma_{1,1}(\pi(t)) :=  \frac{\pi(t) r_0 \mu_0 + (1-\pi(t)) r_1
    \mu_1}{\pi(t) r_0 + (1-\pi(t)) r_1}.
\end{displaymath}
\item if $A(t)=1,$ i.e., arm is played and $Y(t)=1, Z^1(t)=0$ then
%
\begin{displaymath}
 \gamma_{0,1}(\pi(t)) :=  \frac{\pi(t) (1-r_0) \mu_0 + (1-\pi(t))
    (1-r_1) \mu_1}{\pi(t) (1-r_0) + (1-\pi(t)) (1-r_1)}.
\end{displaymath}
%
\item  if $A(t)=1,$ i.e., arm is played and $Y(t)=0, Z^1(t)=1$ then
\begin{displaymath}
\gamma_{1,0}(\pi(t)) := 
\begin{cases}
\pi(t) & \mbox{ for rested bandit,} \\
\gamma_{1,1}(\pi(t)) & \mbox{for restless bandit.}
\end{cases}
\end{displaymath}
\item if $A(t)=1,$ i.e., arm is played and $Y(t)=0, Z^1(t)=0$ then
\begin{displaymath}
\gamma_{0,0}(\pi(t)) := 
\begin{cases}
\pi(t) & \mbox{for rested bandit,} \\
\gamma_{0,1}(\pi(t)) & \mbox{for restless bandit.}
\end{cases}
\end{displaymath}
\item  if $A(t)=0,$ i.e., arm is not played and $Y(t)=1$ then
{\small{
\begin{displaymath}
\Gamma_{1}(\pi(t)) := 
\begin{cases}
\pi(t) & \mbox{for rested bandit,} \\
\pi(t)\mu _0 + (1-\pi(t))\mu _1 & \mbox{for restless bandit.}
\end{cases}
\end{displaymath}
}}
\item if $A(t)=0,$ i.e., arm is not played and $Y(t)=0$ then
\begin{displaymath}
\Gamma_{0}(\pi(t)) := \pi(t).
\end{displaymath}
\end{enumerate}

From \cite{Bertsekas95a}, we know that the $\pi(t)$ captures the
information about the history $H_t$, and it is a sufficient
statistic. It suggests that the optimal policies can be restricted to
stationary Markov policies.  In this, one can obtain the optimum value
function by solving suitable dynamic program, it will be given in
later part of this section.

Let us define the value function under initial action $A_1$ and
availability $Y_1$
\begin{align*}
V_S := {} & \mbox{value function under $A_1=1, Y_1=1$} \\
\widetilde{V}_S := {} & \mbox{value function under $A_1=1, Y_1=0$} \\ 
V_{NS} := {} & \mbox{value function under $A_1=0, Y_1=1$} \\
\widetilde{V}_{NS} := {} & \mbox{value function under $A_1=0, Y_1=0$}
\end{align*}
We can write the following.
{\footnotesize{
\begin{eqnarray}
V_S(\pi) =  \rho(\pi) + \beta [ \rho (\pi )
 \{ {\theta ^1}(\pi ,1) V({\gamma _{1,1}}(\pi )) + 
(1 - {\theta ^1}(\pi ,1)) \widetilde{V}({\gamma _{1,1}}(\pi ))\} \nonumber  \\ 
  + (1 - \rho (\pi )\{ {\theta ^1}(\pi ,1) V({\gamma _{0,1}}(\pi )) + (1 - {\theta ^1}(\pi ,1))
\widetilde{V}({\gamma _{0,1}}(\pi ))\}],
\label{eqn:V-S-pi} 
\end{eqnarray}
}}
{\small{
\begin{eqnarray}
V_{NS}(\pi) = w + \beta [{\theta ^0}(\pi ,1)
V(\Gamma_1(\pi) ) + (1 - {\theta ^0}(\pi ,1))
\widetilde{V}(\Gamma_1(\pi))], 
\label{eqn:V-NS-pi}
\end{eqnarray}
}}
{\footnotesize{
\begin{eqnarray}
\widetilde{V}_S(\pi) =  \xi(\pi) + 
\beta [\rho (\pi )\{ {\theta ^1}(\pi ,0)V({\gamma _{1,0}}(\pi )) 
+ (1 - {\theta ^1}(\pi ,0))
\widetilde{V}({\gamma _{1,0}}(\pi ))\} \nonumber  \\
  + (1 - \rho (\pi )\{ {\theta ^1}(\pi ,0)V({\gamma _{0,0}}(\pi )) 
+ (1 - {\theta ^1}(\pi ,0))
\widetilde{V}({\gamma _{0,0}}(\pi ))\}],
\label{eqn:V-widetilde-S-pi} 
\end{eqnarray}
}}
{\small{
\begin{eqnarray}
\widetilde{V}_{NS}(\pi) =  w + \beta [{\theta ^0}(\pi ,0)V(\Gamma_0(\pi))
+ (1-{\theta ^0}(\pi ,0))\widetilde{V}(\Gamma_0(\pi))].
\label{eqn:V-widetilde-NS-pi}
\end{eqnarray}
}}
Here $\xi(\pi) = \pi \eta_0 + (1 - \pi )\eta_1, \rho (\pi) = \pi r_0 +
(1 - \pi )r_1.$ The optimal value function $V(\pi,y)$, is determined
by solving the following dynamic program
\begin{eqnarray}
V(\pi) = \max \{ {V_S}(\pi),{V_{NS}}(\pi)\}, \nonumber
\\ \widetilde{V}(\pi) = \max \{ {{\widetilde V}_S}(\pi),{{\widetilde
    V}_{NS}}(\pi)\}.
\label{eq:dynamic-program-a}
\end{eqnarray}
These are  dynamic programs for single-armed rested as well as restless 
bandit problems. 
Now, we proceed to present the main results of this work.
\section{Structural results}
\label{sec:struct-results}
We now begin with some of structural results on value functions,
showing convexity and threshold type policy.
\begin{lemma}(Convexity of value function)
\item 
\begin{enumerate}
\item For fixed $w$, $V(\pi),V_S(\pi),V_{NS}(\pi), \widetilde{V}(\pi), \widetilde{V}_{T}(\pi )$ and $\widetilde{V}_{NS}(\pi)$ are  convex functions of $\pi.$
\item For a fixed $\pi$, $V(\pi),V_S(\pi),V_{NS}(\pi),
\widetilde{V}(\pi), \widetilde{V}_{T}(\pi)$ and $\widetilde{V}_{NS}(\pi)$ are non decreasing and convex in $w.$
\end{enumerate}
\label{lemma:convex-pi-w} 
\end{lemma}
A sketch of the proof is in Appendix~\ref{app:lemma-convex-pi-w}.  We
first define a threshold or monotone policy for the single armed
bandit problem and then prove that the optimal policy is of this kind
under some restriction on model parameters.
\begin{definition}(Threshold type policy) 
A policy is said to be threshold type,  if one of the following is true.
\begin{enumerate}
\item The optimal action is to play the arm  $\forall \pi.$
\item The optimal action is to not play the arm $\forall \pi.$
\item There exists a threshold $\pi^*$ such that $\forall \pi \leq \pi^*$
 the optimal action is to play the arm and not to play the arm otherwise.
\end{enumerate}
\label{Def:threshold}
\end{definition}

\subsection{Threshold structure  of optimal policy (case $\mu_0 > \mu_1$)}
\label{sec:Threshold-case1}
The following lemma provides sufficient conditions for monotonicity
of the optimal value function.
\begin{lemma}(Monotone value functions)
If
\begin{enumerate}
\item $0 \leq \eta_{0} < r_{0} < \eta_{1} < r_{1} \leq 1,$ 
\item $\mu_0 > \mu_1,$
\item $\rho_1 > \rho_0,$
\item $\theta^a(\pi,1)>\theta^a(\pi,0),$ and
  $\theta^a(\pi,y)>\theta^a(\pi ',y),$ for $\pi' > \pi,$
\end{enumerate} 
then for $\pi' \geq \pi$ implies $V(\pi) \geq V(\pi ')$ and $\widetilde{V}(\pi) \geq \widetilde{V}(\pi ').$
\label{lemma:monotone-valuef}
\end{lemma}
A sketch of the proof is given in
Appendix~\ref{app:lemma-monotone-valuef}.
\begin{remark}
The lemma says that if the rewards, observation and transition
probabilities follow certain order than the optimal value functions
are monotone with belief $\pi.$ This result can be
utilized to prove that optimal policy is a monotone policy. A monotone policy is one where the actions are monotone over state space.
\end{remark} 

To have monotone optimal policy, we first prove that the difference
between the value functions $V_S(\pi)$ and $V_{NS}(\pi),$ is monotonic
in $\pi.$ Similarly, we prove this for $\widetilde{V}_S(\pi)$ and
$\widetilde{V}_{NS}(\pi).$
\begin{lemma}(Isotone difference property)
For fixed $w$ and conditions of Lemma~\ref{lemma:monotone-valuef} 
\begin{enumerate}
\item $(V_S(\pi) - V_{NS}(\pi))$
 is decreasing in $\pi,$
\item $(\widetilde{V}_S(\pi) - \widetilde{V}_{NS}(\pi))$ 
 is decreasing in $\pi,$  
\end{enumerate}  
\label{lemma:monotone-policy}
\end{lemma}
We describe the proof in Appendix~\ref{app:lemma-monotone-policy}.

Let $S_1:= [0,1] \times \{1\},$ $S_0:= [0,1] \times \{0\},$
$a^*(\pi):= \arg \max \{V_S(\pi),V_{NS}(\pi)\}$ and
$\widetilde{a}^*(\pi):= \arg \max
\{\widetilde{V}_S(\pi),\widetilde{V}_{NS}(\pi)\}.$ Then the following
theorem gives monotone optimal policy on belief $\pi.$
\begin{theorem}(Monotone optimal policy)
\begin{enumerate}
\item If the value function $V:S_1 \times A \rightarrow \mathbb{R}$
  has isotone difference on $S_1\times A$ then there exists a non
  increasing optimal policy $a^*: S_1 \rightarrow A$ on belief $S_1.$
\item If the value function $\widetilde{V}:S_0 \times A \rightarrow
  \mathbb{R}$ has isotone difference on $S_0\times A$ then there
  exists a non increasing optimal policy $\widetilde{a}^*: S_0
  \rightarrow A$ on belief $S_0.$
\end{enumerate}
\end{theorem}
\begin{IEEEproof}
From Lemma~\ref{lemma:convex-pi-w}, the value functions
$V(\pi),\widetilde{V}(\pi)$ are convex and monotone in $\pi.$ From
Lemma~\ref{lemma:monotone-valuef}, $V(\pi),\widetilde{V}(\pi)$ has
isotone difference property. This implies, there exists $a^*(\pi) \in
\{0,1\}$ that is non increasing in $\pi.$ 
\label{theorem:monotone-policy}  
\end{IEEEproof}

\begin{remark}
Here, we observe that the optimal actions are ordered on belief
space. This indeed is a threshold type policy by
Definition~\ref{Def:threshold}. Note that a monotone policy is a
threshold policy for two actions. Thus isotone difference property
implies a threshold policy result.
\end{remark}
\subsection{Threshold structure of optimal policy (case $\mu_0 < \mu_1$)}
\label{sec:Threshold-case2}
For $\mu_0 < \mu_1,$ different proof technique is necessary to To
claim a threshold type optimal policy.  Here, we will assume
$\theta^a(\pi, y) = \theta^a(y),$ i.e. independent of $\pi.$

We first argue that difference between the value functions $V_S(\pi)$
and $V_{NS}(\pi),$ is monotonic in $\pi$ for special cases. Similarly,
difference between $\widetilde{V}_{S}(\pi)$ and
$\widetilde{V}_{NS}(\pi)$ is monotone in $\pi.$
\begin{lemma}
For fixed $w$ and $\beta,$ and $0\leq \mu_1 - \mu_0 \leq \frac{1}{3},$ 
\begin{enumerate}
\item $(V_S(\pi) - V_{NS}(\pi))$
 is decreasing in $\pi,$
\item $(\widetilde{V}_S(\pi) - \widetilde{V}_{NS}(\pi))$ 
 is decreasing in $\pi,$  
\end{enumerate}
\label{lemma:monotone-valuef-2}   
\end{lemma}
We describe sketch of the proof in
Appendix~\ref{app:lemma-monotone-valuef-2}.
\begin{remark}
The proof of this Lemma is different from the earlier
Lemma~\ref{lemma:monotone-valuef} because here we are not assuming
monotonicity of value functions. Instead here we use the Lipschitz
properties of value functions with respect to $\pi,$ i.e., the value
functions, $V(\pi),\widetilde{V}(\pi)$ have following property
\begin{eqnarray}
|V(\pi_1) - V(\pi_2)| &\leq& \kappa |r_1 - r_0| |\pi_1 - \pi_2|, \nonumber \\
|\widetilde{V}(\pi_1) - \widetilde{V}(\pi_2)| &\leq & \kappa |\eta_1 - \eta_0| |\pi_1 - \pi_2|,
\label{eqn:Lipschitz-property}
\end{eqnarray}
where $\kappa = \frac{1}{1-\beta(\mu_1-\mu_0)}.$ It is true for
$0<\mu_1 - \mu_0 \leq 1/3.$ The Lipschitz-property proof is given in
\cite[Appendix, Lemma $5$]{Meshram16}.

\end{remark}

\begin{theorem}
For fixed $w$ and $\beta,$ and $0\leq \mu_1 - \mu_0 \leq \frac{1}{3},$ 
\begin{enumerate}
\item The optimal policy is threshold type for $V_{T}(\pi)$ and
  $V_{NS}(\pi).$ That is, either $V(\pi ) = V_S(\pi)$ for all $\pi \in
  [0,1]$ or $V(\pi ) = V_{NS}(\pi)$ for all $\pi \in [0,1]$ or there
  exists $\pi^*$ such that
\begin{eqnarray*} 
V(\pi) = \begin{cases}
V_S(\pi) & \mbox {  for $\pi \leq  \pi^*,$ }  \\
V_{NS}(\pi) & \mbox {  for $\pi \geq  \pi^*.$ }
\end{cases}
\end{eqnarray*}
 \item The optimal policy is threshold type for
   $\widetilde{V}_{T}(\pi)$ and $\widetilde{V}_{NS}(\pi).$ That is,
   either $\widetilde{V}(\pi ) = \widetilde{V}_S(\pi)$ for all $\pi
   \in [0,1]$ or $\widetilde{V}(\pi ) = \widetilde{V}_{NS}(\pi)$ for
   all $\pi \in [0,1]$ or there exists $\widetilde{\pi}$ such that
\begin{eqnarray*} 
\widetilde{V}(\pi) = \begin{cases}
\widetilde{V}_S(\pi) & \mbox {  for $\pi \leq  \widetilde{\pi},$ }  \\
\widetilde{V}_{NS}(\pi) & \mbox {  for $\pi \geq  \widetilde{\pi}.$ }
\end{cases}
\end{eqnarray*}
\end{enumerate}
\label{thm:threshold-policy2}
\end{theorem}
\begin{remark}
\begin{itemize}
\item The proof of Theorem~\ref{thm:threshold-policy2} is analogous to
  the Theorem~\ref{theorem:monotone-policy}.
\item 
In Section~\ref{sec:nume_result}, we will present few numerical
examples to illustarte a threshold-type policy for general case, where
we do not make any restriction on $\theta$ and model parameters
$\mu$s.

\end{itemize}
\end{remark}
\section{Index policy for single-armed bandit}
\label{sec:sabindex}
Recall that our interest here is to seek an index-type policy.  We now
define indexability of an arm and then its index.  Let
$\mathcal{G}(w)$ be the subset of state space $S = [0,1] \times
\{0,1\}$ in which it is optimal to not play the arm with subsidy $w,$
it is given as follows.
%
\begin{align}
\mathcal{G}(w) := & \{(\pi,y) \in [0,1] \times \{0,1\} : 
\nonumber \\
&
 V_S(\pi) \leq V_{NS}(\pi) ,
 \widetilde{V}_S(\pi) \leq \widetilde{V}_{NS}(\pi)\}.
\label{eq:G_w_set}
\end{align}
%
Using set $\mathcal{G}(w),$ indexability and index are defined as follows.
\begin{definition}
An arm is indexable if $\mathcal{G}(w)$ is increasing in subsidy $w,$ i.e.,  
\begin{displaymath}
w_2 \le w_1 \Rightarrow \mathcal{G}(w_2) \subseteq \mathcal{G}(w_1).
\end{displaymath} 
\end{definition}
\begin{definition}
The index of an indexable arm is defined as 
\begin{equation}
w(\pi,y) := \inf \{w \in \mathbb{R}:(\pi,y) \in \mathcal{G}(w) , \forall (\pi,y) \in S\}.
\label{Index-def} 
\end{equation}
\end{definition}
\begin{remark} \hspace{2cm} 
\begin{itemize}
\item Note that we can rewrite definition of set $\mathcal{G}(w)$ in
  the following way.
\begin{displaymath}
\mathcal{G}(w) = \left\{ [\pi_L,1] \times \{1\}, [\widetilde{\pi}_L,1] \times \{0\}
\right\},
\end{displaymath}
where $\pi_L:= \min \{ \pi \in [0,1] :V_S(\pi) = V_{NS}(\pi)\},$ and
$\widetilde{\pi}_L:= \min \{ \pi \in [0,1] :\widetilde{V}_S(\pi) =
\widetilde{V}_{NS}(\pi)\}.$
\item If the optimal policy is of threshold type, then $\pi^* = \pi_L$
  and $\widetilde{\pi} = \widetilde{\pi}_L.$
\item To claim  indexability, we require to show that as subsidy $w$
  increases, $\pi_L(w)$ and $\widetilde{\pi}_L(w)$ are	 non-increasing
  in $w.$
\item In general, it is difficult to show indexability and obtain
  index because there is difficulty in proving a threshold type
  policy.
\end{itemize}
\end{remark}

 We next show the indexability and compute the closed form expression
 for the index of a single-armed rested bandit.  The proof of index
 computation is along the lines of \cite{Dayanik02}.

\subsection{Rested single-armed bandit}
We further simplify the rested single-armed bandit problem and make
following assumptions on transition probabilities.
\begin{eqnarray*}
P_{ij}(y,a)= 
\begin{cases}
p_{ij} & \mbox{if $y=a=1,$ } \\
\delta _{ij} & \mbox{if $y=0$ or $a=0.$}
\end{cases}
\end{eqnarray*}
where $\delta _{ij}$ equals to 1 if $i=j$ and 0 otherwise.  Also,
$p_{00} = \mu_0,$ and $p_{10} = \mu_1.$ This indicates that state of
the arm changes if arm is available and does not change when arm is
unavailable. Further we assume $\theta^0(\pi,0)=0.$

We now present a few preliminary results which are used to derive the
index. These results make use of the definition of set
$\mathcal{G}(w)$ and obtain value function expressions.
\begin{lemma} \hspace{2cm}
\begin{enumerate}
\item For $(\pi,0) \in S,$ subsidy $w \in \mathbb{R},$ if $(\pi ,0)
  \in \mathcal{G}(w)$ then $\widetilde{V}(\pi,w)= \frac{w}{{1 - \beta
  }}$ with initial state $(\pi ,0).$


\item For $(\pi ,1) \in S,$ subsidy $w \in \mathbb{R}$, if $(\pi ,1)
  \in \mathcal{G}(w)$ then
\end{enumerate}
{\footnotesize{
\begin{dmath}
V(\pi,w) =  \max \left \{E_{\pi,1}^{\phi_0}\left[\sum_{t=1}^{\infty}  \beta^{(t-1)}(w\mathbf{1}_{\{y(t)=1\}}  + R^1(\pi,0)\mathbf{1}_{\{y(t)=0\}})\right],\right. 
 \left.  \frac{w}{{1 - \beta }}\right \}. 
\label{eqn:bound-V}
\end{dmath}
}}
Here $E_{\pi,1}^{\phi_0}$ is the expectation under policy $\phi_0$
that plays the arm when it is unavailable and otherwise keeps it
rested.
\label{lemnma:bound-V}
\end{lemma}
\begin{IEEEproof}
1. State of the arm does not change when arm is unavailable and not
played. Therefore if $(\pi,0) \in \mathcal{G}(w),$ then it is always
optimal to not play the arm and the expected total discounted reward
starting in state $(\pi,0)$ is $\widetilde{V}(\pi,w)= \frac{w}{{1 -
    \beta }}.$

2. If $(\pi,1) \in \mathcal{G}(w),$ then, the arm may visit $(\pi,0)$
state if it goes unavailable in between. Therefore, the arm is in
either $(\pi,1)$ or $(\pi,0)$ state. In this case, two optimal
policies are possible (a) never play the arm, (b) do not play the arm
when it is in state $(\pi,1)$ and play the arm when it is in
$(\pi,0).$ The expected total discounted reward for policy (a) is
$\frac{w}{{1 - \beta }}$ and for policy (b) is given
in~\eqref{eqn:bound-V}.  
\end{IEEEproof}

We now define $E_{\pi,1}^{\phi_1}$ as the expectation under policy
$\phi_1$ that always plays the arm. Then we can evaluate the total
expected discounted reward under $\phi_1$ for initial state $(\pi,1).$
It is
\begin{equation*}
\Psi(\pi,1) :=  E_{\pi,1}^{\phi_1} \left[ \sum_{t=1}^{\infty} \beta^{(t-1)}
R^1(\pi(t), y(t))\right].
\end{equation*}
We can derive lower bound on $\Psi(\pi,1)$ in terms of $\eta_0,$ 
\begin{equation}
\Psi(\pi,1) > \frac{R^1(\pi,0)}{1-\beta} = \frac{\pi \eta_0 + (1-\pi)\eta_1}{1-\beta} > \frac{\eta_0}{1-\beta}.
\label{eq:cond2}
\end{equation}
\begin{lemma}
If subsidy $w$ is smaller than $\eta_0,$ then  set 
$\mathcal{G}(w) = \emptyset.$
\label{lemma:G-emptyset}
\end{lemma}
\begin{IEEEproof}
The proof is by contradiction.  We first consider case for $y =0.$
Suppose that $(\pi,0) \in \mathcal{G}(w),$ hence, $\mathcal{G}(w) \neq
\emptyset.$ Then, from Lemma~\ref{lemnma:bound-V}, we get
$\widetilde{V}(\pi,w) = \frac{w}{1-\beta}.$ We also obtain
$\widetilde{V}(\pi,w) > \frac{w}{1-\beta}$ because $w <\eta_0 <
R^{1}(\pi,0).$ This contradicts our assumption. Hence claim follows.

Now we consider case for $y =1.$ We assume that $(\pi,1) \in
\mathcal{G}(w).$ Then using Lemma~\ref{lemnma:bound-V}, we have
$V(\pi,w) < \frac{R^1(\pi,0)}{1-\beta}$ because $w<R^1(\pi,0).$
Further, we can derive lower bound $V(\pi,w)\geq
\frac{R^1(\pi,0)}{1-\beta}.$ This contradicts the upper bound and
hence our assumption. Thus $\mathcal{G}(w) = \emptyset.$ This
completes the proof. 
\end{IEEEproof}

If subsidy $w$ is higher than $\eta_0,$ then, set $\mathcal{G}(w)$ can
be non-empty. We will provide sufficient condition on subsidy $w$ for
$\mathcal{G}(w)$ to be non-empty.  Also, if set $\mathcal{G}(w)$ is
nonempty then we give lower bound on subsidy $w.$ This is given in the
next Lemma.
\begin{lemma}
 $(\pi,y) \in \mathcal{G}(w)$ if and only if 
\begin{equation}
w \geq (1-\beta)\frac{E_{\pi,y}^{\phi _1}\left[\sum_{t=1}^{\tau
      -1}\beta ^{(t-1)}
    R^1(\pi(t),y(t))\right]}{1-E_{\pi,y}^{\phi_1}[\beta ^{\tau}]}
\label{eq:w-sub-lowerbd}
\end{equation}
for $\tau > 0.$
\label{lemma:G-nonempty}
\end{lemma}
\begin{IEEEproof}
 We first assume that $(\pi,y) \in \mathcal{G}(w).$ We want to prove
 Eqn.~\eqref{eq:w-sub-lowerbd}.  We know from
 Lemma~\ref{lemnma:bound-V} that if $(\pi,0) \in \mathcal{G}(w),$ then
 $\widetilde{V}(\pi,w) = \frac{w}{1-\beta}$ and if $(\pi,1) \in
 \mathcal{G}(w),$ then $V(\pi,w) = \frac{w}{1-\beta}.$ This is true
 for $w \geq R^1(\pi,0).$ This suggests that the optimal action is not
 to play the arm for all time slots.  The optimization problem
 in~\ref{eqn:opt2} reduces to optimal stopping problem, where arm is
 played until stopping time $\tau-1$ and not played since $\tau.$ Thus
 the expected discounted reward is
\begin{equation*}
E_{\pi,y}^{\phi_1}\left[\sum_{t=1}^{\tau -1}\beta ^{(t-1)} R^1(\pi(t),y(t)) 
 + \sum_{t=\tau}^{\infty} \beta^{t}w \right]
\end{equation*}
This expected reward is upper bounded by $\frac{w}{1-\beta}$ because
not playing arm is always optimal for $(\pi,y) \in \mathcal{G}(w)$ as
shown earlier. Hence
 \begin{equation}
\frac{w}{1-\beta} \geq E_{\pi,y}^{\phi_1}\left[\sum_{t=1}^{\tau -1}\beta ^{(t-1)} R^1(\pi(t),y(t)) 
 + \sum_{t=\tau}^{\infty} \beta^{t}w \right].
 \label{eq:w-bdd}
\end{equation}

We assume that $w$ is lower bounded and Eqn.~\eqref{eq:w-sub-lowerbd}
holds true. Then, it is easy to verify that $(\pi,y) \in
\mathcal{G}(w).$ To see this, make use of the optimal stopping time
policy and Eqn.~\eqref{eq:w-bdd}. 
\end{IEEEproof}
\begin{theorem}
The arm is indexable and  index $w(\pi,y)$ is 
\begin{align}
w(\pi,y) := (1-\beta) \sup_{\tau \in S} \frac{E_{\pi,y}^{\phi _1}\left[\sum_{t=1}^{\tau -1}\beta ^{(t-1)} R^1(\pi(t),y(t))\right]}{1-E_{\pi,y}^{\phi_1}[\beta ^{\tau}]}
\end{align}
Here, $\tau$ is optimal stopping time, it is time until which arm is
played.
\end{theorem}
\begin{IEEEproof}
Note that Eqn.~\eqref{eq:w-sub-lowerbd} is true for every stopping
time $\tau >0.$ That implies not playing the arm is optimal. Further,
the following is true.
\begin{equation}
w \geq (1-\beta) \sup_{\tau \in S} \frac{E_{\pi,y}^{\phi _1}\left[\sum_{t=1}^{\tau -1}\beta ^{(t-1)} R^1(\pi(t),y(t))\right]}{1-E_{\pi,y}^{\phi_1}[\beta ^{\tau}]}.
\label{eq:index}
\end{equation}
In order to show indexability, we need to prove that $\mathcal{G}(w)$
set is monotone in $w.$ From Lemma~\ref{lemma:G-emptyset}, we know
that there is $w$ for which set $\mathcal{G}(w)$ is empty. As $w$
increases this set becomes non-empty. This is clear from
Lemma~\ref{lemma:G-nonempty}.  As subsidy $w$ increases,
Eqn.~\eqref{eq:w-sub-lowerbd} continues to hold for larger subset of
$S=[0,1] \times \{0,1\}.$ Thus indexability holds true by definition
and index can be computed using \eqref{eq:index}. 
\end{IEEEproof}
\section{Numerical Results}
\label{sec:nume_result}

We first present few numerical examples to illustrate threshold type
optimal policy for a restless single-armed bandit. We later demonstrate
the performance of our index policy for rested multi-armed bandit.

\subsection{Examples for a threshold type result}

\begin{figure}
  \begin{center}
    \begin{tabular}{cccc}
      \includegraphics[scale=0.2]{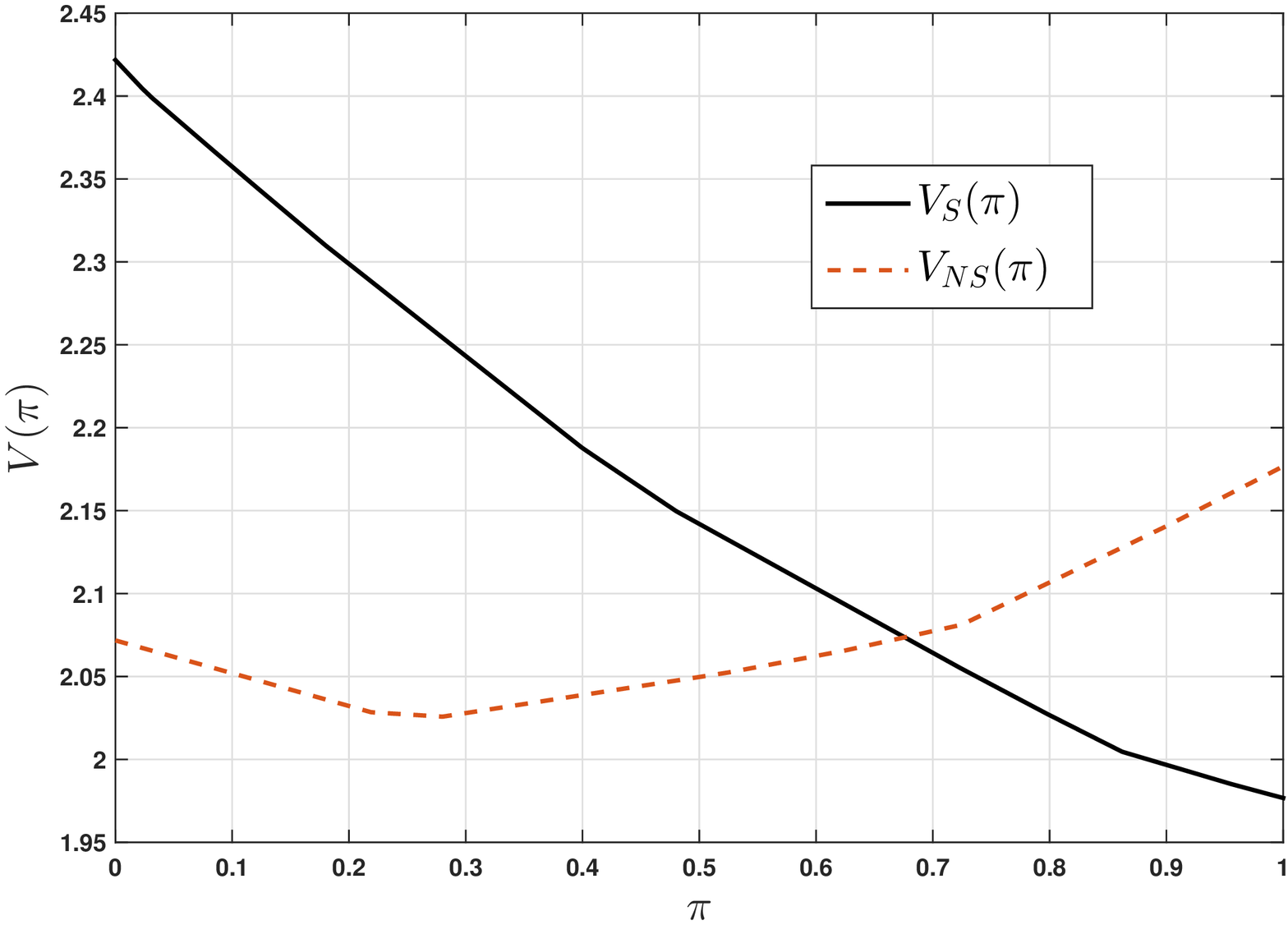}
      & 
      \hspace{-0.6cm} \includegraphics[scale=0.2]{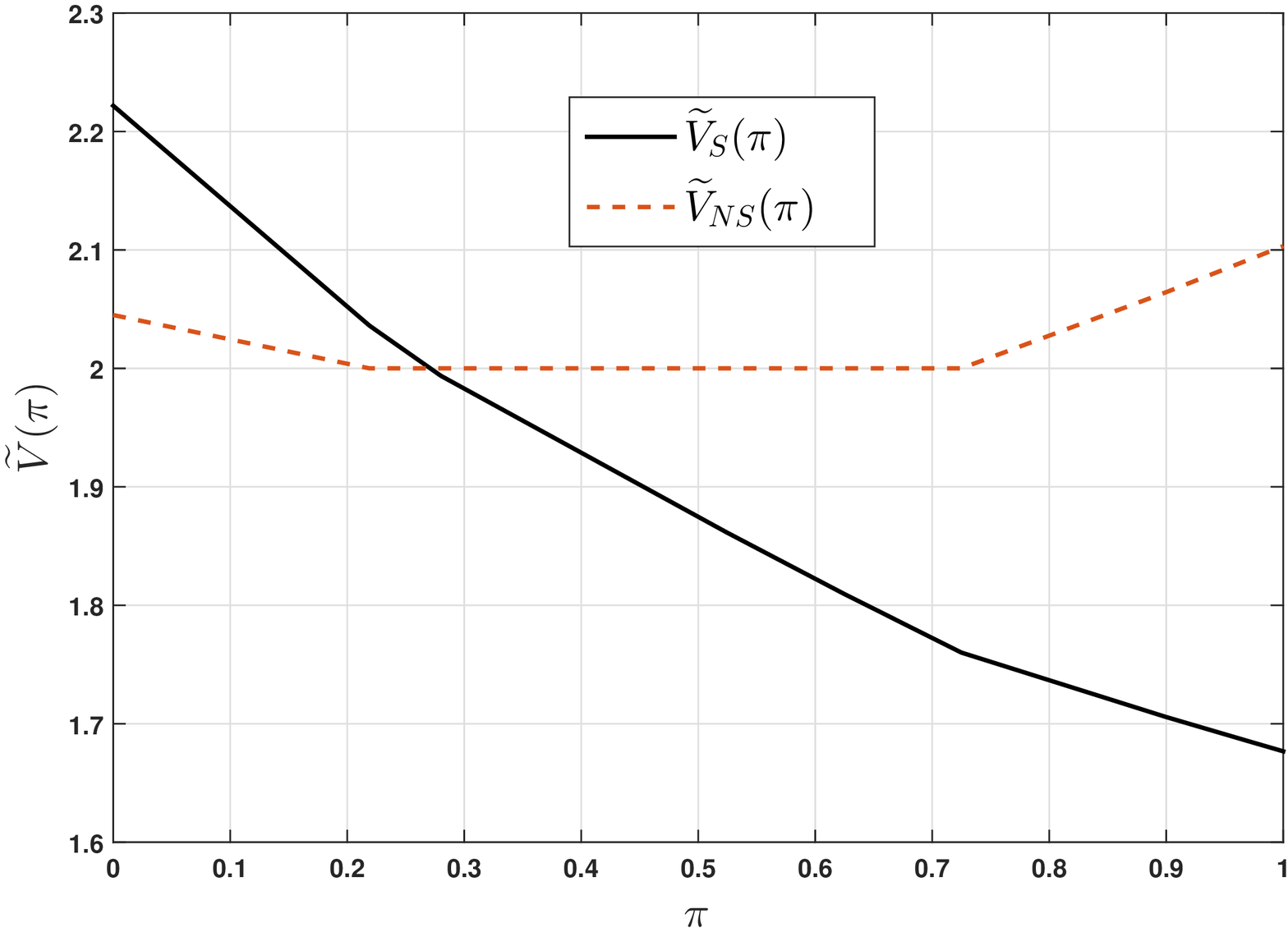} \\ a)
             {\small{$V_S(\pi)$ and $V_{NS}(\pi)$ }} 
             & \hspace{-0.8cm}
             b) {\small{ $\widetilde{V}_S(\pi)$ and $\widetilde{V}_{NS}(\pi)$  }}
     
    \end{tabular}
  \end{center}
  \caption{ In a) $V_S(\pi)$ and $V_{NS}(\pi)$ plotted as function of
    $\pi$ and b) $\widetilde{V}_S(\pi)$ and $\widetilde{V}_{NS}(\pi)$
    plotted as function of $\pi.$ This is plotted for a single-armed
    restless bandit.}
  \label{fig:threshold-Y0-Y1}
\end{figure}

To demonstrate the threshold type result for a single-armed bandit, we use the
following parameters.  $\mu_0 = 0.1, \mu_1 = 0.9, r_0 = 0.4, \eta_0 =
0.1, r_1 = 0.95, \eta_1 = 0.65, \theta^a(\pi,y) = 0.5$ for any $\pi
\in [0,1],$ $a,y \in \{0,1\},$ and $\beta = 0.7.$

In Fig.~\ref{fig:threshold-Y0-Y1}-a), we plot $V_S(\pi)$ and
$V_{NS}(\pi)$ as function of $\pi.$ Similarly, in
Fig.~\ref{fig:threshold-Y0-Y1}-b), we plot value functions
$\widetilde{V}_S(\pi)$ and $\widetilde{V}_{NS}(\pi).$ These plots
suggest that the optimal policy is of a threshold type.

In this case, we have $\mu_1 - \mu_0 = 0.8.$ But to prove analytically
a threshold policy result, we have assumed $0<\mu_1 - \mu_0 < 1/3,$
see Section~\ref{sec:Threshold-case2}. This is a limitation from
analysis because it is very difficult to evaluate closed form
expressions for value functions or introduce monotonicity of value
functions.
%
%

%
\subsection{Performance of index policy}

We now present few numerical examples to illustrate the performance of
index policy and compare this with that of myopic policy. This is
done for rested single-armed bandit. Note that this is different from
standard rested bandits because here arms are available
probabilistically in each time slot.
Recall that in an index policy, the arm with highest index is
played in given time slot. In myopic policy, the arm with highest
immediate expected reward is played at each time slot. 

We consider number of arms, $N = 5$ and use the following set of
parameters in all examples.
\begin{eqnarray*}
\mu_0 = [0.1,0.9,0.3,0.9,0.3],
\mu_1 = [0.9,0.1,0.9,0.3,0.9], \\
r_0 = [0.2,0.3,0.25,0.4,0.35], 
r_1  =  [0.9,0.95,0.8,0.9,0.6], \\
\eta_0 = [0.1,0.2,0.15,0.3,0.25],
\eta_1 = [0.6,0.65,0.5,0.6,0.3]. 
\end{eqnarray*}
We also set $\rho_0 = r_0,$ $\rho_1 = r_1,$ initial belief and
availability vector of arms is
\[\pi(1) = [0.2,0.4,0.3,0.7,0.5] , y(1) = [1,1,1,1,1].\]

We further have two sets of examples, in first set of examples we
assume that the probability of availability is identical
for all the arms, i.e., $\theta_n^a(\pi,y) = \theta^a(\pi,y).$ In
second set of examples, each arm has different probability of
availability.

\subsubsection{Arms with identical probability of availability}
Here, $\theta_n^a(\pi,y) = \theta^a(\pi,y).$ But we assumed different
reward and transition probabilities. We consider four examples as
given below.
\begin{enumerate}
\item $\theta^1(\pi,1) = 1, \theta^1(\pi,0) = 0$ and $\theta^0(\pi,1)
  = 1$

\item $\theta^1(\pi,1) = 0.8, \theta^1(\pi,0) = 0$ and
  $\theta^0(\pi,1) = 0.7$
\item $\theta^1(\pi,1) = 0.8, \theta^1(\pi,0) = 0.4$ and
  $\theta^0(\pi,1) = 0.7.$
\item $\theta^1(\pi,1) = 0.35, \theta^1(\pi,0) = 0.75$ and
  $\theta^0(\pi,1) = 0.9.$
\end{enumerate}
From value function equations
\eqref{eqn:V-S-pi}--\eqref{eqn:V-widetilde-NS-pi}, we can observe the
influence of $V(\pi)$ and $\widetilde{V}(\pi)$ on each other, that is
based on different value of $\theta^a(\pi,y).$

\begin{table}[ht]
\centering
\caption{We use $\theta^1(\pi,1) = 1, \theta^1(\pi,0) = 0$ and
  $\theta^0(\pi,1) = 1$}
    \scalebox{0.85}{
\begin{tabular}{|c|c|c|c|}
\hline
\multicolumn{4}{|c|}{Total discounted  cumulative reward} \\
\hline
$\beta$ & Myopic policy & Index policy & $\%$ Gain in  \\
& & & index policy  \\ \hline
0.95 & 15 & 17 & 13.33\\ \hline
 0.8 & 3 & 3.18 & 6.3\\ \hline
 0.6 & 1.9 & 1.8 & -4.2\\
 \hline
\end{tabular}}
\label{table:1}
\end{table}

The first example captures the scenario, where there is no influence of
$V(\pi)$ and $\widetilde{V}(\pi)$ on each other. In
Tables~\ref{table:1}, we show a detailed comparison of discounted
cumulative reward using index based policy and myopic policy. Also, we
observe that the index policy performs better than myopic policy for
large values of discount parameters $\beta,$ i.e., $\beta$ closer to
$1.$ In this example, myopic policy gives better peformance over index
policy for $\beta = 0.6.$

\begin{table}[ht]
\centering
\caption{We set $\theta^1(\pi,1) = 0.8, \theta^1(\pi,0) = 0$ and
  $\theta^0(\pi,1) = 0.7$}
    \scalebox{0.85}{
\begin{tabular}{|c|c|c|c|}
\hline
\multicolumn{4}{|c|}{Total discounted  cumulative reward} \\
\hline
$\beta$ & Myopic policy & Index policy & \% Gain in  \\ 
& & & index policy \\ \hline
 0.95& 8.33 & 10 & 20\\ \hline
 0.8 & 1.97 & 2.5 & 26.9\\ \hline
 0.6 & 1 & 1.35 & 25\\
 \hline
\end{tabular}}
\label{table:3}
\end{table}
In our second example, we consider $\theta^1(\pi,0) = 0,$ i.e., no
influence from $V(\pi)$ on $\widetilde{V}(\pi)$ but $\theta^1(\pi,1) =
0.8,$ and $\theta^0(\pi,1) = 0.7,$ i.e., there is influence from
$\widetilde{V}(\pi)$ on $V(\pi),$ see
Eqn.~\eqref{eqn:V-widetilde-S-pi}. The performance is given in
Table~\ref{table:3}. It suggests that the index policy yields up to
$20 \%$ gain in discounted cumulative reward compared to myopic
policy. In this example, index policy gives better performance compared
to myopic policy even for $\beta =0.6.$

\begin{table}[ht]
\centering
\caption{We use $\theta^1(\pi,1) = 0.8, \theta^1(\pi,0) = 0.4$ and
  $\theta^0(\pi,1) = 0.7.$}
  \scalebox{0.85}{
\begin{tabular}{|c|c|c|c|}
\hline
\multicolumn{4}{|c|}{Total discounted  cumulative reward} \\
\hline
$\beta$ & Myopic policy & Index policy & \% Gain in  \\ 
& & & index policy \\ \hline
 0.95 & 13.4  & 15 & 11.94\\ \hline
 0.8 & 3.5 & 3.56 & 1.71\\ \hline
 0.6 & 1.82 & 1.74 & -4.12\\
 \hline
\end{tabular}}
\label{table:4}
\end{table}

In third example, we use $\theta^1(\pi,0) = 0.4,$ $\theta^1(\pi,1) =
0.8,$ and $\theta^0(\pi,1) = 0.7.$ The performance is illustrated in
Table~\ref{table:4}. This example captures a scenario with some
influence from $V(\pi)$ and $\widetilde{V}(\pi)$ on each other.  We
notice that index policy provides gain in cumulative discounted reward
compared to myopic polic for $\beta =0.8, 0.95.$ The index policy
yields up to $12\%$ gain in discounted reward over myopic policy for
$\beta = 0.95.$ But it does not provide any gain for $\beta = 0.6.$

In above first $3$ examples we considered
$\theta^1(\pi,1)>\theta^1(\pi,0),$ see
Table~\ref{table:1}---\ref{table:4}. This implies that the probability
that the arm is available in next slot given that it is not available
and played in current time slot is smaller that the probability of
availability in next slot given the arm is available and played. On
the other hand we consider example of
$\theta^1(\pi,1)<\theta^1(\pi,0)$ in Table~\ref{table:5}, which means
playing an arm when it is not available leads to better chance of it
being available in the next slot than playing when it is available. we
observe similar performance to that of example $3$.


\begin{table}[h]
\centering
\caption{We use $\theta^1(\pi,1) = 0.35, \theta^1(\pi,0) = 0.75$ and
  $\theta^0(\pi,1) = 0.9.$}
  \scalebox{0.85}{
\begin{tabular}{|c|c|c|c|}
\hline
\multicolumn{4}{|c|}{Total discounted  cumulative reward} \\
\hline
$\beta$ & Myopic Policy & Index Policy & \% Gain in  \\
& & &  index policy \\ \hline
 0.95& 12.3 & 13.82 &12.35\\ \hline
 0.8 & 3.3 & 3.25& -1.3\\ \hline
 0.6 & 1.7 & 1.6 & -5.88\\
 \hline
\end{tabular}}
\label{table:5}
\end{table}
\subsubsection{Arms with non identical probability of availability}
In next set of examples we have considered the scenario where arms
have same rewards and transition probabilities but different
probabilities of availability.  The transition probabilities are,
$\mu_0 = 0.9, \mu_1 = 0.3$ and rewards, $\eta_0 = 0.1, \eta_1 = 0.6$
and $r_0 = 0.2, r_1=0.9.$ The initial belief and availability vector
for arms are \[\pi(1) = [0.2,0.4,0.3,0.7,0.5] , y(1) = [1,0,1,0,1].\]
Example illustrating two possible scenarios were considered with
parameters shown in Table~\ref{table:multiarm with diff theta}.
\begin{table}[h]
\centering
\caption{Second set of examples - probabilities of availability}
\label{table:multiarm with diff theta}
\scalebox{.85}{
\begin{tabular}{|l|c|c|c|c|c|c|c|c|c|c|}
\hline
Arm               & \multicolumn{2}{c|}{1} & \multicolumn{2}{c|}{2} & \multicolumn{2}{c|}{3} & \multicolumn{2}{c|}{4} & \multicolumn{2}{c|}{5} \\ \hline
Example               & 1          & 2         & 1          & 2         & 1          & 2         & 1          & 2         & 1         & 2          \\ \hline
$\theta^1(\pi,1)$ & 0.5        & 0.5       & 0.5        & 0.3       & 0.8        & 0.8       & 0.5        & 0.5       & 1         & 1          \\ 
$\theta^1(\pi,0)$ & 0.7        & 0.7       & 0.5        & 0.5       & 0.9        & 0.9       & 0.5        & 0.5       & 0         & 0.2        \\ 
$\theta^0(\pi,1)$ & 0.9        & 0.9       & 0.5        & 0.6       & 0.7        & 0.7       & 0.5        & 0.5       & 1         & 1          \\ \hline
\end{tabular}}
\end{table}
From Table~\ref{table:Different availability Compare} we can see that
index policy performs better compared to myopic policy. The index
policy gives upto $16$ to $18\%$ gain over myopic policy. The authors
observed that, in both examples, myopic policy chose arms 1,3 and 5 in
initial time slots and later on kept choosing arm 5. The index policy
chose arm 5 from the beginning. This again suggests the
``far-sightedness'' of the index policy in accounting for future
states and availability of arms.


\begin{table}[h]
\centering
\caption{Arms with different availability probability}
  \scalebox{0.85}{
\begin{tabular}{|c|c|c|c|}
\hline
\multicolumn{4}{|c|}{Total discounted  cumulative reward} \\
\hline
Example & Myopic Policy & Index Policy & \% Gain in  \\
& & & index policy \\ \hline
 1& 54.12 & 64 & 18.2\\ \hline
 2 & 54 & 63& 16.6\\ 
 \hline
\end{tabular}}
\label{table:Different availability Compare}
\end{table}

\section{Concluding remarks}
\label{disc}
In this paper we presented monotonicity results and showed that the
optimal policy is of threshold type under some model
restrictions. Though this is generally true, it is difficult to prove
without any restriction on model parameters. We have demonstrated this
via numerical examples. Hidden states and interdependence between
$V(\pi)$ and $\widetilde{V}(\pi)$ makes it difficult to get closed
form expression for the threshold.

For a rested single-armed bandit with availability constraints, we
have shown that the arm is indexable and derived a formula for
index. The index can also be calculated by the value iteration
algorithm. From numerical examples, we observed that index policy
performs better than myopic policy for some cases. This suggests that,
index policy accounts for the future availability of arms and hence
gives better performance.  In future we seek to obtain some numerical
scheme to compute the index for restless bandits with constrained
arms.

\bibliography{varun}

\section*{Appendix}


\subsection{Proof of Lemma~\ref{lemma:convex-pi-w}}
\label{app:lemma-convex-pi-w}
1. In this part, We  prove $V(\pi)$ is convex by induction and use that to show other value functions are also convex.

Let 
\begin{align*}
V_1(\pi) =  \max \{\pi r_0 + (1-\pi)r_1 , w\}
\end{align*}
\begin{dmath*}
V_{n+1,S}(\pi) = \rho(\pi) + \beta [\rho (\pi )\{ {\theta ^1}(\pi ,1)V_n({\gamma _{1,1}}(\pi)) +  (1 - {\theta ^1}(\pi ,1))\widetilde{V}_n({\gamma _{1,1}}(\pi))\} 
 + (1 - \rho (\pi )\{ {\theta ^1}(\pi ,1)V_n({\gamma _{0,1}}(\pi)) + (1 - {\theta ^1}(\pi ,1))\widetilde{V}_n({\gamma _{0,1}}(\pi))\}]
\end{dmath*}
\begin{dmath*}
V_{n+1,NS}(\pi) = w + \beta [{\theta ^0}(\pi ,1)V_n(\Gamma_1(\pi)) + (1 - {\theta ^0}(\pi ,1))\widetilde{V}_n(\Gamma_1(\pi))]
\end{dmath*}
\begin{align}
V_{n+1}(\pi) =  \max \{V_{n+1,S}(\pi) , V_{n+1,NS}(\pi)\}
\label{eqn:max of Vs and Vns}
\end{align}
Now define
\begin{align*}
b_0 :=  & [\pi {\mu _0}(1 - {r_0}) + (1 - \pi ){\mu _1}(1 - {r_1}), \\
         & \pi (1 - {\mu _0})(1 - {r_0}) + (1 - \pi )(1 - {\mu _1})(1 - {r_1})]^T \\
b_1 := & [\pi {\mu _0}{r_0} + (1 - \pi ){\mu _1}{r_1}, \\
         & \pi (1 - {\mu _0}){r_0} + (1 - \pi )(1 - {\mu _1}){r_1}]^T \\   
\hat{b}_0 = & \theta^1(\pi,1)b_0 \\ 
\hat{b}_1 = & \theta^1(\pi,1)b_1                       
\end{align*}
clearly, $V_1(\pi)$ is linear and hence convex. If $V_n(\pi),\widetilde{V}_n(\pi)$ is convex in $\pi$ then we can write 
{\footnotesize{
\begin{dmath*}
V_{n+1,S}(\pi) =  \left\| b_1 \right\|_1 + \beta  \left\| \hat{b}_1 \right\|_1 V_n\left({\frac{\hat{b}_1}{\left\| \hat{b}_1 \right\|_1}}\right) + \beta  \left\| \hat{b}_1 \right\|_1 \widetilde{V}_n \left({\frac{\hat{b}_1}{\left\| \hat{b}_1 \right\|_1}}\right) + 
  \beta \left\| \hat{b}_0\right\|_1 V_n\left({\frac{\hat{b}_0}{\left\| \hat{b}_0 \right\|_1}}\right) + \beta \left\| \hat{b}_0 \right\|_1 \widetilde{V}_n\left({\frac{\hat{b}_0}{\left\| \hat{b}_0 \right\|_1}}\right).
\end{dmath*} }}

From \cite{Astrom69}[Lemma~$2$], we can argue that $V_{n+1,S}(\pi)$ is convex in $\pi.$ Similarly,  we can show this for other value functions.

2. In this part, We can rewrite \eqref{eqn:max of Vs and Vns}, in form of $V_{n+1,S}(\pi,w)$ and $V_{n+1,NS}(\pi,w)$ as function of $w$. We can see that $V_1(\pi,w)$ is monotone non decreasing and convex in $w$. $V_{n+1,S}(\pi,w)$ is constant plus a convex sum of four non decreasing convex function of $w.$ $V_{n+1,NS}(\pi,w)$ is the sum of three non decreasing function of $w.$ The convexity is preserved under max operation so $V_{n+1}(\pi,w)$ is also non decreasing and convex in $w$ and using induction, all $V_n(\pi,w)$ follows the same. As $V_n(\pi,w) \rightarrow V(\pi,w)$ and this complete the proof for $V(\pi)$. Similarly, we can show this for other value functions.
\subsection{Proof of Lemma~\ref{lemma:monotone-valuef}}
\label{app:lemma-monotone-valuef}
The proof can be done via induction technique. The basic intuition behind ordering rewards, transition and observation probabilities on belief $\pi$ is to get monotone decreasing value functions over $\pi.$ \\  

Assume that $V_n(\pi)$ and $\widetilde{V}_n(\pi)$ is non increasing in $\pi.$ Lets take $\pi' \geq \pi$ and playing an arm is optimal. Then induction step
\begin{dmath*}
V_{n+1}(\pi) = \rho(\pi) + \beta [ \rho (\pi )
 \{ {\theta ^1}(\pi ,1) V_n({\gamma _{1,1}}(\pi )) + 
(1 - {\theta ^1}(\pi ,1)) \widetilde{V}_n({\gamma _{1,1}}(\pi ))\}  \\ 
  + (1 - \rho (\pi )\{ {\theta ^1}(\pi ,1) V_n({\gamma _{0,1}}(\pi )) + (1 - {\theta ^1}(\pi ,1))
\widetilde{V}_n({\gamma _{0,1}}(\pi ))\}]
\end{dmath*}
Here $\rho(\pi)$ is decresing in $\pi,$ i.e. $\rho(\pi')<\rho(\pi)$ for $\pi'>\pi.$ Hence
\begin{dmath*}
V_{n+1}(\pi) \geq \rho(\pi') + \beta [ \rho (\pi )
 \{ {\theta ^1}(\pi ,1) V_n({\gamma _{1,1}}(\pi )) + 
(1 - {\theta ^1}(\pi ,1)) \widetilde{V}_n({\gamma _{1,1}}(\pi ))\}  \\ 
  + (1 - \rho (\pi )\{ {\theta ^1}(\pi ,1) V_n({\gamma _{0,1}}(\pi )) + (1 - {\theta ^1}(\pi ,1))
\widetilde{V}_n({\gamma _{0,1}}(\pi ))\}]
\end{dmath*}
From our assumptions $\mu_0>\mu_1, \rho_1>\rho_0$ and $\theta^a(\pi,y)>\theta^a(\pi',y),$ we get stochastic ordering on obervation and availability probability, i.e., $[\rho(\pi), 1-\rho(\pi)]^T \leq_s [\rho(\pi'), 1-\rho(\pi')]^T$ and $[\theta^a(\pi,y), 1-\theta^a(\pi,y)]^T \leq_s [\theta^a(\pi',y), 1-\theta^a(\pi',y)]^T.$ Then
\begin{dmath*}
V_{n+1}(\pi) \geq \rho(\pi') + \beta [ \rho (\pi' )
 \{ {{\theta^'}^1}(\pi ,1) V_n({\gamma _{1,1}}(\pi )) + 
(1 - {\theta ^1}(\pi ,1)) \widetilde{V}_n({\gamma _{1,1}}(\pi ))\}  \\ 
  + (1 - \rho' (\pi )\{ {{\theta^'}^1}(\pi ,1) V_n({\gamma _{0,1}}(\pi )) + (1 - {{\theta^'}^1}(\pi ,1))
\widetilde{V}_n({\gamma _{0,1}}(\pi ))\}]
\end{dmath*}
Now $\gamma _{1,1}(\pi),\gamma _{0,1}(\pi)$ are increasing in $\pi$ and $V_n(\pi),\widetilde{V}(\pi)$ are decreasing in $\pi,$ then we have
\begin{dmath*}
V_{n+1}(\pi) \geq \rho(\pi') + \beta [ \rho (\pi' )
 \{ {\theta ^1}(\pi' ,1) V_n({\gamma _{1,1}}(\pi' )) + 
(1 - {\theta ^1}(\pi' ,1)) \widetilde{V}_n({\gamma _{1,1}}(\pi' ))\}  \\ 
  + (1 - \rho (\pi' )\{ {\theta ^1}(\pi' ,1) V_n({\gamma _{0,1}}(\pi' )) + (1 - {\theta ^1}(\pi' ,1))
\widetilde{V}_n({\gamma _{0,1}}(\pi' ))\}]
\end{dmath*}
\begin{dmath*}
V_{n+1}(\pi) \geq V_{n+1}(\pi').
\end{dmath*}
Similarly we can show that $\widetilde{V}_{n+1}(\pi) \geq \widetilde{V}_{n+1}(\pi').$ This is true for every $n \geq 1.$ From Chapter~$7$ of \cite{Bertsekas95a} and Proposition~$2.1$ of Chapter~$2$ of \cite{Bertsekas95b}, $V_n(\pi) \rightarrow V(\pi),$ uniformly and similarly $\widetilde{V}_n(\pi) \rightarrow \widetilde{V}(\pi).$ Hence $V(\pi) \geq V(\pi')$ and $\widetilde{V}(\pi) \geq \widetilde{V}(\pi')$ for $\pi' \geq \pi.$ 
\subsection{Proof of Lemma~\ref{lemma:monotone-policy}}
\label{app:lemma-monotone-policy}
From Lemma~\ref{lemma:monotone-valuef} $V_S(\pi)$ is strictly decreasing in $\pi$ and $V_{NS}(\pi)$ is nonincreasing in $\pi.$ 

Let $f(\pi) = V_S(\pi) -V_{NS}(\pi)$ and $f(\pi)$ is decreasing in $\pi,$ i.e $f(\pi)<f(\pi')$ for $\pi>\pi'.$ This implies that we need to show 
\begin{dmath}
V_{S}(\pi)-V_{NS}(\pi) < V_{S}(\pi')-V_{NS}(\pi')
\label{eqn:submodular-I}
\end{dmath}
Rearranging~\ref{eqn:submodular-I} we need to show 
\begin{dmath}
V_{S}(\pi)-V_{S}(\pi') < V_{NS}(\pi)-V_{NS}(\pi') 
\label{eqn:submodular-II}
\end{dmath}
Rested bandit: Right hand side of ~\eqref{eqn:submodular-II} is $0.$ We know $V_S(\pi)$ is decreasing, hence our claim follows.\\
Restless bandit: When $\rho_0=0,\rho_1=1$ similar argument holds and claim follows. But in other cases, the claim holds under some restrictions on $\beta$ and to prove this one required to use Lipschitz properties~\eqref{eqn:Lipschitz-property} of value functions.


\subsection{Proof of Lemma~\ref{lemma:monotone-valuef-2}}
\label{app:lemma-monotone-valuef-2}
As before $f(\pi) = V_S(\pi)-V_{NS}(\pi).$ In order to prove that $f(\pi)$ is decreasing, we need to show that its partial derivative w.r.t. $\pi$ is negative. 

Taking partial derivative of $f(\pi)$ w.r.t. $\pi,$ we obtain 
\begin{dmath}
\frac{\partial f(\pi)}{\partial \pi} = \frac{\partial V_S(\pi)}{\partial \pi} - \frac{\partial V_{NS}(\pi)}{\partial \pi}
\end{dmath}

Next using Lipschitz property of value function~\ref{eqn:Lipschitz-property}, we can obtain following upper bound  on the sampling value function
\begin{dmath*}
\frac{\partial V_S(\pi)}{\partial \pi} \leq (\rho_1-\rho_0) \kappa \{-1+2\beta(\mu_1-\mu_0)\},
\end{dmath*}
and lower bound on non sampling value function 
\begin{dmath*}
 \frac{\partial V_{NS}(\pi)}{\partial \pi}   \geq -\kappa (\rho_1-\rho_0)|\mu_1-\mu_0| .
\end{dmath*}
Hence 
\begin{dmath}
\frac{\partial f(\pi)}{\partial \pi}  \leq (\rho_1-\rho_0)\kappa \{-1+2\beta(\mu_1-\mu_0) + \beta|\mu_1-\mu_0|\}
\end{dmath}
We want $\{-1+2\beta(\mu_1-\mu_0) + \beta|\mu_1-\mu_0|\} < 0$ for the derivative of $f(\pi)$ to be negative. This holds true when $0<\mu_1-\mu_0<\frac{1}{3}.$ 

It is possible that $V_S(\pi),V_{NS}(\pi)$ is not differential w.r.t $\pi.$ In that case right partial derivative should be taken. Such partial derivative exists because $V_S(\pi),V_{NS}(\pi)$ are convex and bounded.  
\end{document}